# Type III events, flares and CMEs, in the extremely active period October-November 2003

E. Mitsakou[1], M. Thanassa[1], A. Hillaris[1], P. Preka-Papadema[1], X. Moussas[1], A. Caroubalos[2], C. Alissandrakis[3], P. Tsitsipis[4], A. Kontogeorgos[4], J-L. Bougeret[5], G. Dumas[5]

[1] *Section of Astrophysics, Astronomy and Mechanics, Department of Physics, University of Athens*
[2] *Department of Informatics, University of Athens*
[3] *Section of Astro-Geophysics, Department of Physics, University of Ioannina*
[4] *Department of Electronics, Technological Education Institute of Lamia*
[5] *Observatoire de Paris, Departement de Recherche Spatiale*

**Abstract.** The type III radio bursts trace the propagation of populations of energetic electrons through the solar corona which, more often than not, are associated with energy release on the Sun. A dataset of type III radio bursts observed in the range 20-650 MHz during the period of extraordinary solar activity, from 20 October to 5 November 2003, by the ARTEMIS-IV radio spectrograph was compiled and analyzed. Their parameters were compared with characteristics of associated flares (Ha and GOES SXR) and CMEs reported in the NGDC, the SPIDR and the LASCO archives, respectively. In this report, we examine the correlation between energetic particles and major manifestations of solar activity such as flares and CMEs.

**Keywords:** solar radio type III bursts, Ha flares, GOES SXR, CMEs, ARTEMIS-IV

**PACS:** 96.60.-j, 96.60.Q-

## INTRODUCTION

We study the period of intense solar activity from 20 October to 5 November 2003. This specific period has already attracted considerable attention because of its unexpected behaviour ([1],[2]). As it is well known, the maximum of the solar cycle 23 lasted from 2000 to 2002. That means that the year 2003 is on the descending phase of the solar cycle. However, during the period October-November 2003, an extreme solar activity occurred, which significantly affected the interplanetary medium as well as the terrestrial environment. The solar activity for the above mentioned period originates mainly in active regions 484, 486, and 488 which have been extremely active from their first appearance until their disappearance beyond the west limb. There are indications of "triggering" among the flares occurring in the three active regions. Prominent among the numerous events recorded in the period that we study is the X28/3B flare, which occurred on 4 November 2003, one of the more powerful flares ever detected.

# METRIC TYPE III RADIO BURSTS

In this report, we focus our attention on the type III radio events which have been recorded by ARTEMIS-IV during the period from 20 October to 5 November 2003. We study the association of this type III radio activity with energetic events, such as Ha and SXR flares and CMEs. ARTEMIS-IV is the solar radio spectrograph of the University of Athens, operating at the Thermopylae Station since 1996 ([3]). Observations cover the frequency range from 20 to 650 MHz, in the metric wavelengths. Radio emissions from the Sun at greater wavelengths, decametric and kilometric, cannot be observed from the ground, as they do not penetrate the Earth's ionosphere. Emissions at metric wavelengths originate close to the Sun, while decametric, hectometric and kilometric emissions originate at greater distances and in the interplanetary medium. Thus, the ARTEMIS-IV radio spectrograph can detect the coronal radio emission.

Type III bursts are usually interpreted as electron beams propagating through the corona and in the interplanetary medium ([4] and the references within). In the dynamic and differential spectrum, they appear as almost vertical features, drifting rapidly in frequency as time progresses, while the propagation of the electrons occurs along open magnetic fields. However, in the inner corona, different types of bursts often occur, such as the U or the J bursts, produced by electron beams propagating along closed or curved magnetic field lines, or even reverse drift bursts, due to electron propagating towards the Sun. Extensive literature exists on these types of radio bursts ([5]).

During the period from 20 October to 5 November 2003, the ARTEMIS-IV radio spectrograph has recorded all of the above mentioned type III radio bursts. In our study, we examine their temporal association with Ha and GOES SXR flares, using a time window of 1 minute before and after the type III burst. The SXR flare data were obtained from the GOES X-ray Sensor, while the Ha flares from the National Geophysical Data Center. Furthermore, we investigate whether a coronal mass ejection was recorded half an hour before and after the burst. We also make a spatial correlation with the specific CMEs. For that purpose we have used the SOHO LASCO CME Catalog.

The total number of events that were studied was 206 type III radio bursts. These bursts were separated into four categories, depending on their association with other energetic events. The first category (FIGURE 1) consists of type III radio bursts that are well correlated with flares and CMEs, that is to say 5 events, although in 3 cases the CME was not spatially collated with the flare. In the second category (FIGURE 2), we have 22 bursts that were correlated only with flares, but were without an associated CME. The third category (FIGURE 3) includes 5 bursts that were correlated with CMEs, but not with a flare. Finally, in the fourth category (FIGURE 4) 173 type III radio bursts were neither associated with a flare, nor with a CME. Nevertheless, in 111 of these bursts, considerable SXR enhancements have been observed, which were not recorded as flares in the NGDC archives.

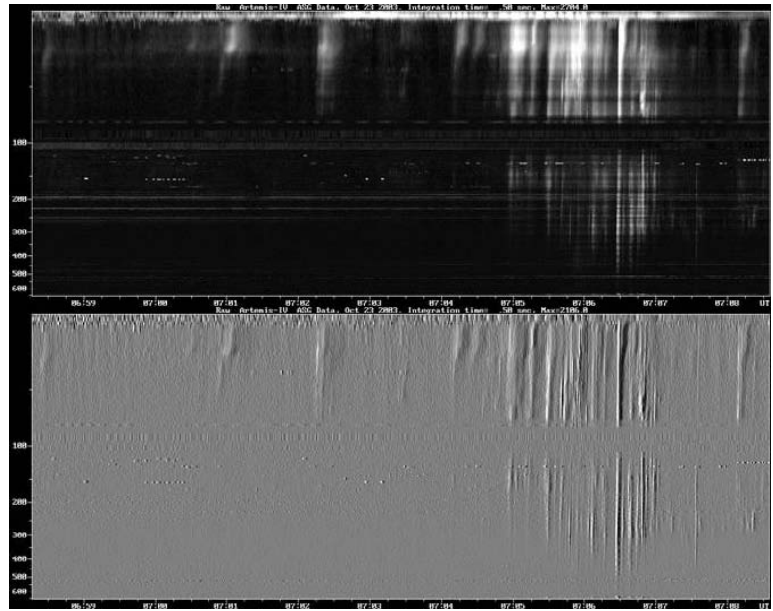

**FIGURE 1.** The dynamical and the differential spectrum of the type III radio event on the 23 October 2003, which has been recorded by the ARTEMIS-IV radio spectrograph at approximately 07:05 UT. This event has been associated with an M flare, as well as with a CME, and it is therefore listed in the first category of our classification.

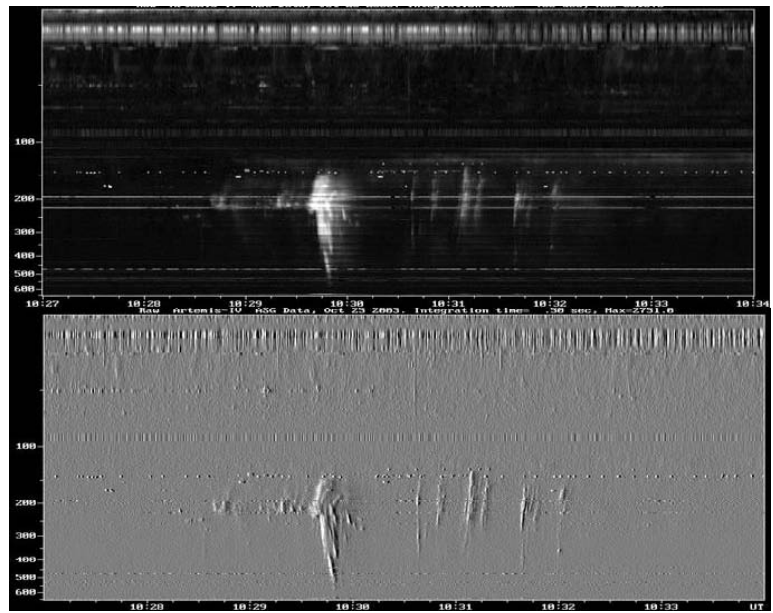

**FIGURE 2.** The dynamical and the differential spectrum of the type III radio event on the 25 October 2003, which have been recorded by the ARTEMIS-IV radio spectrograph at approximately 10:30 UT. This event has been associated with an M flare, but on with a CME, and it is therefore listed in the second category of our classification.

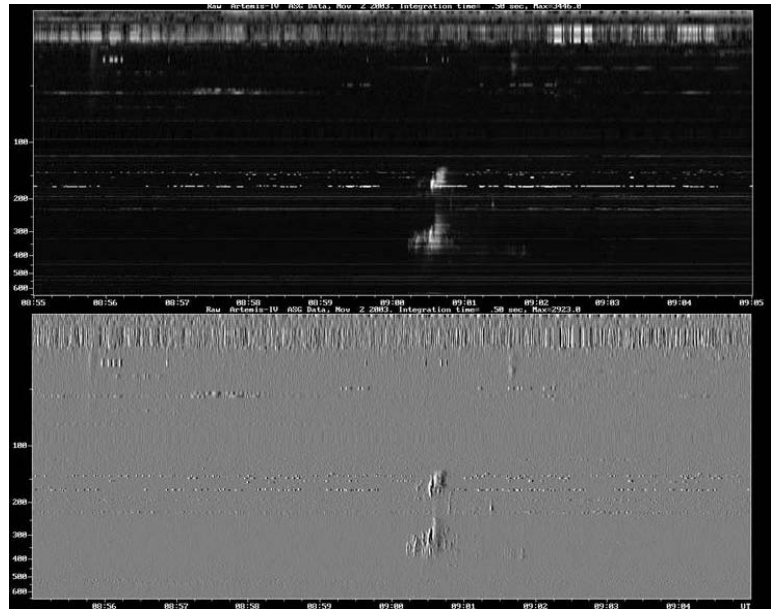

**FIGURE 3.** The dynamical and the differential spectrum of the type III radio event on the 2 November 2003, which has been recorded by the ARTEMIS-IV radio spectrograph at approximately 09:00 UT. This event has been associated with a Halo CME, but not with a flare, and is therefore listed in the third category of our classification.

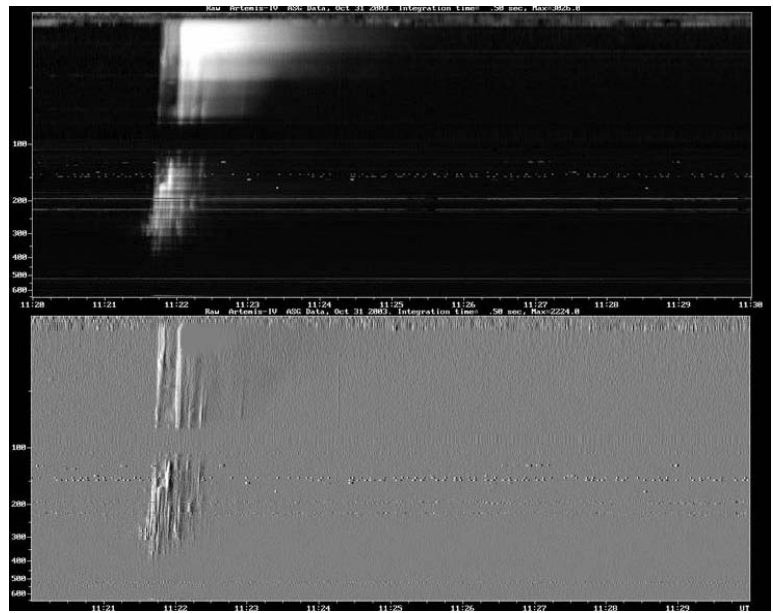

**FIGURE 4.** The dynamical and the differential spectrum of the type III radio event on the 31 October 2003, which has been recorded by the ARTEMIS-IV radio spectrograph at approximately 11:20 UT. This event has neither been associated with a flare, nor with a CME, and is therefore listed in the fourth category of our classification.

# CONCLUSIONS

206 type III radio bursts were recorded by ARTEMIS-IV during the period from 20 October to 5 November 2003, a period of extreme solar activity. In this study, the temporal association of the type III radio bursts with solar energetic events (flares and CMEs) was examined. The radio events were classified into 4 categories, summarized as follows:

a. Category 1 (type III with flare and CME). 5 events are closely temporally associated with a flare and a CME, though in 3 of them the CME and flare are not spatially correlated.

b. Category 2 (type III with flare). 22 bursts are closely associated with a flare, but not with a CME.

c. Category 3 (type III with CME). 5 type III radio events are closely associated with a CME, but not with a flare.

d. Category 4 (type III without flare or CME). For 173 type III radio events neither a flare, nor a CME was recorded in the time window introduced for the association in our study.

Further investigation is carried out in order to determine the exact correlation between the events, as well as the physical process that lies beneath it.